\documentclass[apl,reprint,superscriptaddress]{revtex4-1}

\usepackage[dvipdfmx]{graphicx}
\usepackage{dcolumn}
\usepackage{bm}
\usepackage{color}

\begin{document}
\title{Phase-resolved Spin-Wave Tomography}

\author{Yusuke Hashimoto}
\affiliation{Advanced Institute for Materials Research, Tohoku University, Sendai 980-8577, Japan}

\author{Tom H. Johansen}
\affiliation{Department of Physics, University of Oslo, 0316 Oslo, Norway}
\affiliation{Institute for Superconducting and Electronic Materials, University of Wollongong, Northfields Avenue, Wollongong, NSW 2522, Australia}

\author{Eiji Saitoh}
\affiliation{Advanced Institute for Materials Research, Tohoku University, Sendai 980-8577, Japan}
\affiliation{Institute for Materials Research, Tohoku University, Sendai 980-8577, Japan}
\affiliation{Advanced Science Research Center, Japan Atomic Energy Agency, Tokai 319-1195, Japan}

\date{\today}

\begin{abstract}
The propagation dynamics of spin waves are represented by their dispersion relations.
Recently, we have developed a method, called spin-wave tomography (SWaT), to obtain dispersion relation of spin waves in the long wavelength regime, so-called pure magnetostatic waves. 
In our previous studies on SWaT, phase information of spin waves was disregarded.
In this report, we demonstrate an advanced SWaT analysis, called phase-resolved spin-wave tomography (PSWaT), to realize the direct observation of the amplitude and the phase of spin waves.
The PSWaT spectra are obtained by separating the real and the imaginary components of the complex Fourier transform in the SWaT analysis.
We demonstrate the PSWaT spectra of spin waves excited by the photo-induced demagnetization in a Bi-doped garnet film, reflecting the characteristic features of the complex dynamical susceptibility affected by magnetostatic coupling in the film.
\end{abstract}

\pacs{63.20.kk, 75.30.Ds, 75.40.Gb, 75.78.Jp}


\maketitle

A spin wave is a collective excitation of spin precessions in coupled spin systems.
A phase shift between one spin to another is determined by the wavevector of the spin wave~\cite{Schneider2008,Lee:2008kv,Stancil:2009ux,Serga2010}.
In spintronics and magnonics devices, data may be transferred by the phase of spin waves~\cite{Chumak2015}, so that its manipulation would be one of cornerstones for the development of spin-wave based applications~\cite{Khitun2008}.

The phase-resolved observation of spin waves has been realized with a Mach-Zehnder interferometer~\cite{Kostylev:2005fu,Schneider2008}, phase-sensitive Raman light scattering~\cite{Fallarino:2013ex,Fohr:2009gz,Serga:2006gc}, and time-resolved magneto-optical (TRMO) spectroscopy~\cite{Crooker:1996iy,Kirilyuk2010,Carpene:2010gs,Lenk2011,Bossini2015,Janusonis2016,Bossini2017}.
Nowadays, the phase-resolved imaging of the propagation dynamics of optically-excited spin waves with sub-picosecond time resolution and micrometer spatial resolution has been realized with a TRMO imaging method~\cite{Satoh2012,Au:2013kb,Hashimoto2014,Yoshimine:2014hq,Ogawa2015,Iihama2016,Hashimoto:2017jb,Hashimoto:2017tu,Kamimaki2017,Savochkin:2017ei}.

Recently, we have developed spin-wave tomography (SWaT): reconstruction of dispersion relation of spin waves~\cite{Hashimoto:2017jb}.
The SWaT is based on a convolution theory, a Fourier transform, and a pump-and-probe magneto-optical imaging method for spin-wave propagation.
In SWaT, the phase information of spin waves was lost due to the limitations in the data analysis procedure.

In this study, we demonstrate phase-resolved SWaT: reconstruction of dispersion relation of spin waves with their phase information.
The phase of spin waves is determined by an advanced analysis procedure separating the real and the imaginary components of the complex Fourier transform in the SWaT analysis.
The phase of the PSWaT spectra represents the complex dynamical susceptibility ($\mbox{\boldmath $\chi$}$) and the spatial symmetry of spin waves.
Moreover, PSWaT provides information about the temporal evolution and the orientation of the torque generating spin waves, exerted on the local magnetization (${\bf M}$).
Therefore, 
PSWaT is a very powerful method to investigate the excitation and the propagation dynamics of spin waves.



\begin{figure}
\includegraphics[width=7cm]{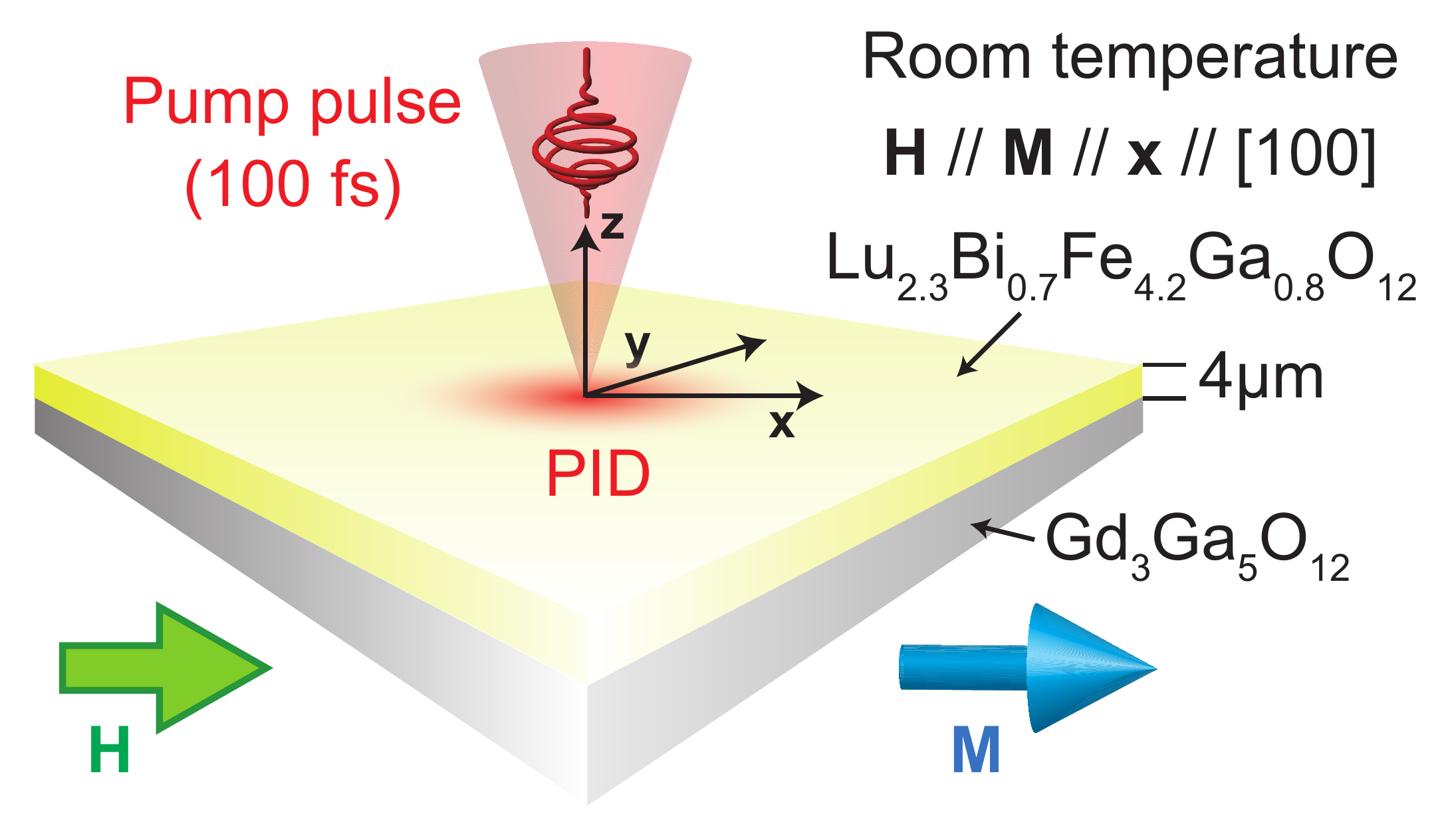}
\caption{\label{fig:ExeConf} 
A schematic illustration of the experimental configuration and the rectangular coordinate ($x$, $y$, $z$) used in this study.
The sample surface is in the $x$-$y$ plane with the $x$-axis along the orientation of ${\bf M}$ parallel to the [100] axis.
The orientation of ${\bf M}$ was controlled by an external magnetic field, ${\bf H}$.
The direction normal to the sample surface is along the $z$-axis.
The pump pulse is focused on the sample surface at the origin of the coordinate and induces a photo-induced demagnetization (PID), which is the source of the spin waves investigated in this study.
}
\end{figure}

The rectangular coordinate ($x$, $y$, $z$) shown schematically in Fig.~\ref{fig:ExeConf} is used in this report.
The sample surface is in the $x$-$y$ plane with the $x$-axis parallel to the orientation of $\bf M$.
The $z$-axis is along the sample normal.
In our experiments, the pump and probe pulses illuminate the sample along the $z$-axis.
The propagation dynamics of optically-excited spin waves in the $x$-$y$ plane ($m({\bf r}, t)$) are observed with a time-resolved magneto-optical imaging system~\cite{Zvezdin:1997ub}.

We consider spin waves in an in-plane magnetized garnet film excited by a photo-induced effective field ${\bf h}(${\bf r}$, t) = h_{z}(${\bf r}$, t) + ih_{y}(${\bf r}$, t)$, where $h_{z}$ and $h_{y}$ are the fields along the $z$- and $y$-axes, respectively.
The sample magnetization was controlled by applying an in-plane magnetic field, ${\bf H}$, with the intensity $H_{0}$.
We assume $H_{0} \gg h_{x}, h_{y}$ so that the precession angle of $\bf M$ is small.
We consider a magnetic medium that has a translational invariance in the $x$-$y$ plane.
Then, the propagation dynamics of spin waves are represented by a complex dynamical susceptibility $\mbox{\boldmath $\chi$}$(${\bf r}, t$), or its Fourier transform (FT) $\mbox{\boldmath $\chi$}$(${\bf k}, \omega$)~\cite{White:2013cw}.



The PSWaT spectra are obtained by analyzing the propagation dynamics of the spin waves ${\bf m}({\bf r}, t)$ excited by a photo-induced torque ${\bf h}({\bf r}, t)$.
Then, ${\bf{m}}({\bf r}, t)$ is given by
\begin{eqnarray}
\label{eq:m}
{\bf{m}}({\bf r}, t) = \int\int\int \mbox{\boldmath $\chi$}({\bf r} - {\bf a}, t - \tau){\bf h}({\bf a}, {\bf \tau})d{\bf a}d{\bf \tau}.
\end{eqnarray}
According to the convolution theory, the FT of ${\bf{m}}({\bf r}, t)$ is written as
\begin{eqnarray}
\label{eq:M1}
{\bf{m}}({\bf k}, \omega) &=& \int\int\int {\bf{m}}({\bf r}, t) \exp^{-i({\bf{k}}\cdot {\bf{r}} - \omega t)}d{\bf r}dt\\
\label{eq:M2}
&=&\mbox{\boldmath $\chi$}({\bf k}, \omega){\bf h}({\bf k}, \omega),
\end{eqnarray}
where ${\bf m}$(${\bf k}$, $\omega$) and ${\bf h}$(${\bf k}$, $\omega$) are the FT of ${\bf m}$(${\bf r}$, t) and ${\bf h}$(${\bf r}$, t), respectively.
When the $z$-component of ${\bf m}$ ($m_{z}$) is observed through the Faraday effect, the SWaT intensity spectrum $|m_{z}(${\bf k}$, \omega)|$ is obtained~\cite{Hashimoto:2017jb}.
On the other hand, PSWaT described in the present paper deals with the phase information of spin waves by separating the real and the imaginary components of the FT spectra in the calculation of $m_{z}({\bf k}, \omega)$ by Eq.~\ref{eq:M1}.

\begin{figure}
\includegraphics[width=8cm]{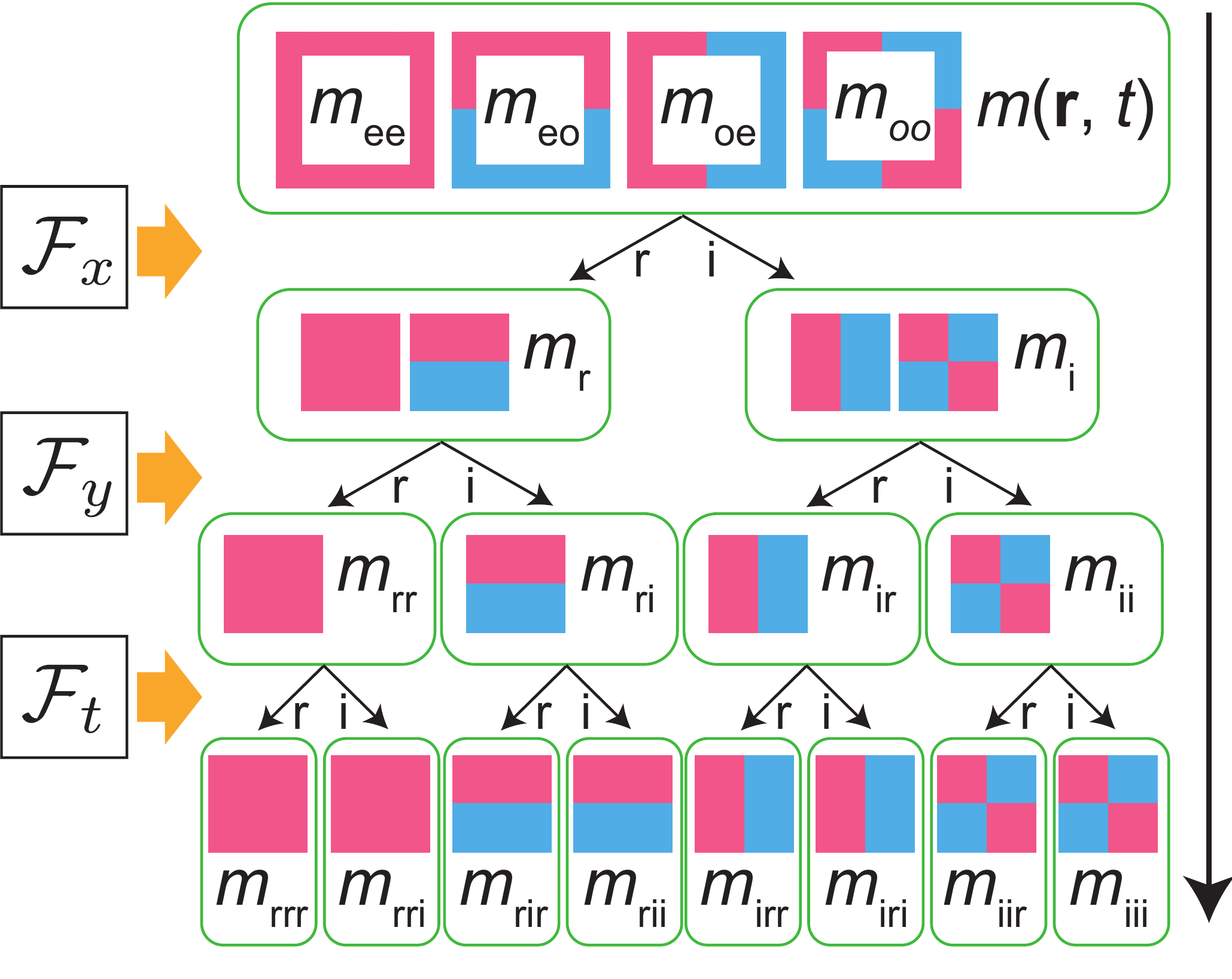}
\caption{\label{FigSetUp} 
The PSWaT spectra are obtained by analyzing $m({\bf r}, t)$ data with the three-dimensional FT along the $x$-axis ($\mathcal{F}_{x}$), the $y$-axis ($\mathcal{F}_{y}$), and the time-axis ($\mathcal{F}_{t}$).
The real (r) and imaginary (i) components of the complex FT spectra are separated.
}
\end{figure}

The PSWaT spectra are obtained by applying a time-space FT to the experimentally observed $m({\bf r}, t)$, as shown schematically in Fig.~\ref{FigSetUp}.
We here define the PSWaT spectra $m_{pqs}$, where $p$, $q$, and $s$ label the real ($r$) and the imaginary ($i$) components of the FT along the $x$-, $y$-, and time axes, respectively.

Let us first consider the spatial FT in the PSWaT analysis.
For simplicity, we start from the one-dimensional FT for $m$($x$) with real values.
Any one-dimensional data $m(x)$ can be decomposed into even ($m_{e}(x)$) and odd ($m_{o}(x)$) components. Thus, we write
\begin{eqnarray}
\label{eq:symmetryX}
m(x) = m_{e}(x) + m_{o}(x).
\end{eqnarray}
Then, the FT of $m(x)$ along the $x$-axis ($\mathcal{F}_{x}$) is given by
\begin{eqnarray}
\label{eq:FTsym}
\mathcal{F}_{x}\{m(x)\} = m(k_{x}) = m_{r}(k_{x}) + m_{i}(k_{x})
\end{eqnarray}
with 
\begin{eqnarray}
m_{r}(k_{x}) &= \mathcal{R}\{m(k_{x})\} &= \mathcal{F}_{x}\{m_{e}(x)\},\\
m_{i}(k_{x}) &= \mathcal{I}\{m(k_{x})\} &= \mathcal{F}_{x}\{m_{o}(x)\},
\end{eqnarray}
where $\mathcal{R}$ and $\mathcal{I}$ are the real and the imaginary components, respectively~\cite{Gray:2012tq}.
Similarly, any two-dimensional data $m(x,y)$ can be decomposed into four components with different symmetries as
\begin{eqnarray}
\label{eq:symmetryXY}
m = m_{ee} + m_{eo} + m_{oe} + m_{oo}.
\end{eqnarray}
$m_{ee}$ ($m_{oo}$) is a component with even (odd) symmetry along both $x$- and $y$-axes, while $m_{eo}$ ($m_{oe}$) is a component with even (odd) symmetry along the $x$-($y$-)axis and odd (even) symmetry along the $y$-($x$-)axis.
The notation of (${\bf r}$) is omitted for clarity.
Then, the FT of $m$ along the $x$-$y$ plane ($\mathcal{F}_{xy}$) gives
\begin{eqnarray}
\label{eq:FT2Dsym}
\mathcal{F}_{xy}\{m({\bf r})\} = m_{rr}({\bf k}) + m_{ri}({\bf k}) + m_{ir}({\bf k}) + m_{ii}({\bf k})
\end{eqnarray}
with $m_{rr}({\bf k}) = \mathcal{F}_{xy}\{m_{ee}({\bf r})\}$, $m_{ri}({\bf k}) = \mathcal{F}_{xy}\{m_{eo}({\bf r})\}$, $m_{ir}({\bf k}) = \mathcal{F}_{xy}\{m_{oe}({\bf r})\}$, and $m_{ii}({\bf k}) = \mathcal{F}_{xy}\{m_{oo}({\bf r})\}$.
Therefore, the complex components obtained by the spatial FT represent the spatial symmetry of the observed spin waves.

The PSWaT spectra are obtained by applying a temporal FT ($\mathcal{F}_{t}$) to the four components of the $m_{rr}$, $m_{ir}$, $m_{ri}$, and $m_{ii}$.
We write 
\begin{eqnarray}
\label{eq:FTXYrrr}
m_{pqr}({\bf k}, \omega) &=& \mathcal{R}\{\mathcal{F}_{t}\{m_{pq}({\bf k}, t)\}\},\\
\label{eq:FTXYrri}
m_{pqi}({\bf k}, \omega) &=& \mathcal{I}\{\mathcal{F}_{t}\{m_{pq}({\bf k}, t)\}\}.
\end{eqnarray}
We define the phase of the PSWaT spectra for the temporal FT as $\phi_{pq}({\bf k}, \omega)$ = $\tan^{-1}[m_{pqi}({\bf k}, \omega)/m_{pqr}({\bf k}, \omega)]$.
This represents the initial phase of spin waves by $m({\bf k}, \omega) \propto \exp\{i\phi_{pq}({\bf k}, \omega)\}$.
This is the basic concept of PSWaT.

We examine SWaT and PSWaT for a 4-$\mu$m thick Lu$_{2.3}$Bi$_{0.7}$Fe$_{4.2}$Ga$_{0.8}$O$_{12}$ film grown on a Gd$_{3}$Ga$_{5}$O$_{12}$ (001) substrate.
The sample was in-plane magnetized with a saturation magnetization of 780 Oe, which was measured with a vibrating sample magnetometer.
The orientation of ${\bf M}$ was controlled by applying an external magnetic field of 560 Oe along the [100] axis .

The SWaT and the PSWaT spectra are obtained by analyzing the propagation dynamics of optically-excited spin waves, observed with a time-resolved magneto-optical imaging method~\cite{Hashimoto2014}.
We used a pulsed light source with 800-nm center wavelength, 100-fs time duration, and a repetition frequency of 1 kHz.
This beam was divided into two beams: pump and probe.
The circularly-polarized pump beam was focused on the sample surface by an objective with a radius of 2 $\mu$m.
The excited spin waves were observed with the probe beam, which was weakly focused on the sample surface with a radius of roughly 1 mm.
The wavelength of the probe beam was changed with an optical parametric amplifier to 630 nm, where the sample shows a large Faraday rotation angle (5.2 degree) and high transmissivity (41 $\%$).
The fluence of the pump and the probe beams were 1.2 J cm$^{-2}$ and 0.2 mJ cm$^{-2}$, respectively.
The propagation dynamics of spin waves were observed with a magneto-optical imaging system based on a rotating analyzer method using a CCD camera~\cite{Hashimoto2014}.
The spatial resolution of the obtained images is one micrometer, determined by the diffraction limit of the probe beam.
All the experiments were performed at room temperature.
The details of the experimental setup were summarized in Ref.~\onlinecite{Hashimoto2014}.

\begin{figure*}
\includegraphics[width=17cm]{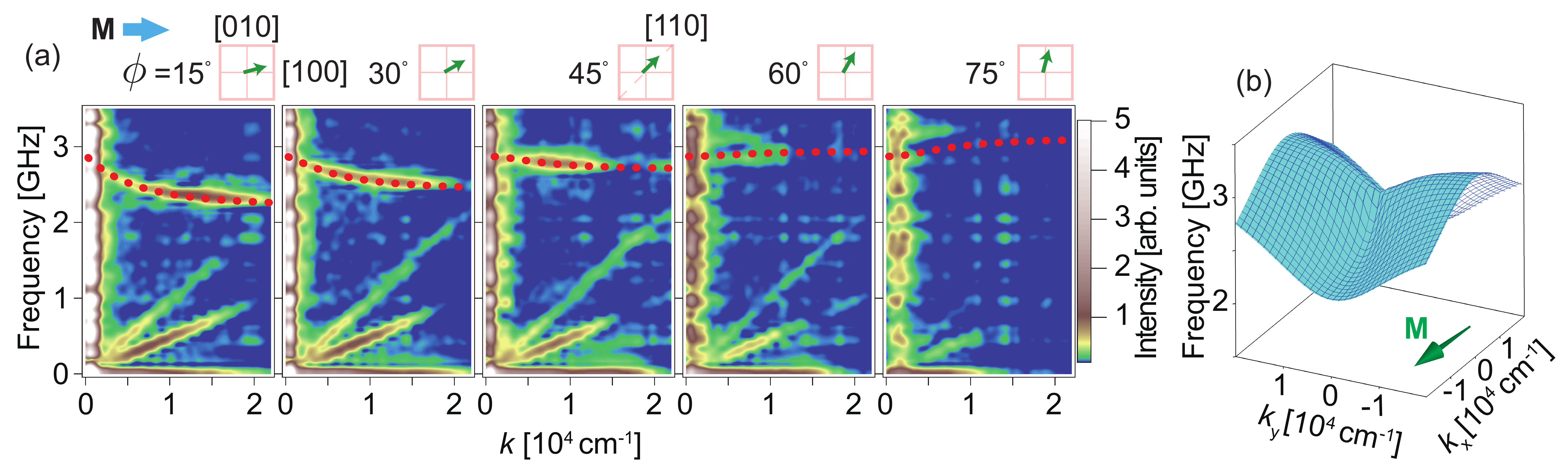}
\caption{\label{FigSWaT} 
(a) The angular dependence of the SWaT spectra for various ${\bf k}$ directions with an angle ($\phi$) between ${\bf M}$ and ${\bf k}$.
The red dotted lines are the dispersion relation of the volume mode of magnetostatic waves calculated with the Damon-Eshbach theory~\cite{Hurben:1995fb}.
(b) The dispersion relation of the volume mode of magnetostatic waves calculated with the parameters obtained in Ref.~\cite{Hashimoto:2017jb}.
The direction of $\bf M$ is indicated by the green arrow.
}
\end{figure*}

Let us first show the SWaT spectra under the in-plane magnetic field of 560 Oe, where the spin-wave excitation has been attributed to photo-induced demagnetization (PID)~\cite{Hashimoto:2017jb}.
An illumination of a magnetic medium with an intense pump pulse decreases local magnetization by PID~\cite{Hashimoto2014}.
The detailed mechanism of PID is out of the scope of the present study.
This change in local magnetization ($\delta {\bf m}({\bf r})$) induces an extra demagnetization field (${\bf h}^{\rm PID}({\bf r})$) in the surrounding area and generates spin waves~\cite{vanKampen:2002jc,Lenk2011,Au:2013kb}.
The demagnetization field lasts for a long time so that we write ${\bf h}^{\rm PID}({\bf r},t) = {\bf h}^{\rm PID}({\bf r})\Theta(t)$, where $\Theta(t)$ is the Heaviside step function.
Thus, we obtain
\begin{eqnarray}
\label{eq:FTPID}
{\bf m}^{\rm PID}({\bf k},\omega) = -\frac{{\bf h}^{\rm PID}({\bf k})}{\omega} \mbox{\boldmath $\chi$}({\bf k},\omega), 
\end{eqnarray}
for $\omega \neq 0$.
The SWaT spectra along various ${\bf k}$ directions are shown in Fig.~\ref{FigSWaT}(a)~\cite{Hashimoto:2017jb}.
We see branches caused by spin waves in the frequency range from 2 GHz to 3 GHz.
A dispersion relation of spin waves is shown by the white dashed lines in Fig.~\ref{FigSWaT}(a), which are the fitting results with the Damon-Eshbach theory~\cite{Hurben:1995fb} representing the dispersion relation of magnetostatic waves.
By using the parameters obtained in Ref.~\cite{Hashimoto:2017jb}, the dispersion relation of magnetostatic waves is reconstructed as shown in Fig.~\ref{FigSWaT}(b).
The strong angular dependence in the dispersion relation for magnetoelastic waves is due to the anisotropic nature of the dipolar coupling~\cite{Hurben:1995fb,Stancil:2009ux,Hashimoto:2017jb}.

\begin{figure*}
\includegraphics[width=17cm]{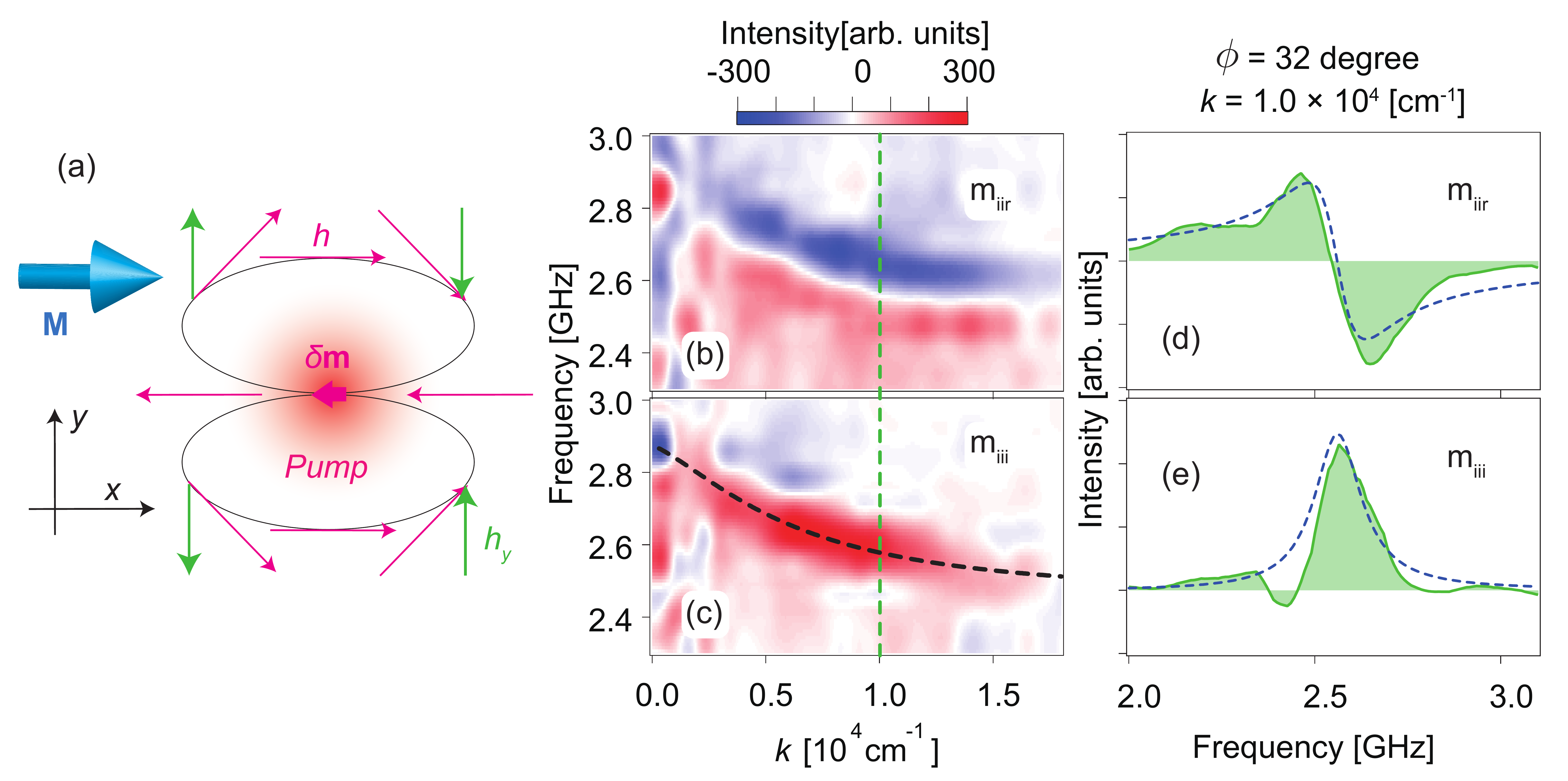}
\caption{\label{SWaTAt560Oe} 
(a) A schematic illustration of the field induced by photo-induced demagnetization ($\delta {\bf m}$).
The field along the $y$-axis ($h_{y}$) generates spin waves with mirror symmetries about both the $x$- and $y$-axes.
The $m_{iir}$ (b) and $m_{iii}$ (c) components of the PSWaT spectra along the direction of $\phi$ = 32 degree.
An in-plane magnetic field of 560 Oe was applied along the [100] axis.
The black dashed line in (c) is the dispersion relation of magnetostatic waves calculated with the Damon-Eshbach theory~\cite{Hurben:1995fb}.
Figures (d) and (e) are the cross-sections of the data shown in (b) and (c) along at $k$ = 1.0 $\times$ 10$^{4}$ cm$^{-1}$ (green dashed lines in (b) and (c)).
The dashed lines in (d) and (e) are calculated with Eqs.~\ref{eq:LLGR} and \ref{eq:LLGI}, respectively, with $\omega_{0}({\bf k})/2\pi$ = 2.56 GHz and $\alpha$ = 0.03.
}
\end{figure*}

Next, we show the PSWaT spectra in Fig.~\ref{SWaTAt560Oe}.
Since ${\bf h}^{\rm PID}({\bf r})$ has mirror symmetry for both $x$ and $y$ axes as schematically drawn in Fig.~\ref{SWaTAt560Oe}(a), the excited spin waves appear in the $m_{ii}$ component.
Figures~\ref{SWaTAt560Oe}(b) and~\ref{SWaTAt560Oe}(c) show the $m_{iir}$ and $m_{iii}$ components of the PSWaT spectra along the direction of $\phi$ = 32 degree, where $\phi$ is the angle between ${\bf M}$ and ${\bf k}$.
We see branches with different features in the $m_{iir}$ and $m_{iii}$ components.
Figures~\ref{SWaTAt560Oe}(d) and~\ref{SWaTAt560Oe}(e) show the cross-sections of the PSWaT spectra shown in Figs.~\ref{SWaTAt560Oe}(b) and~\ref{SWaTAt560Oe}(c), respectively, along at $k = 1.0 \times 10^{4}$ rad cm$^{-1}$.
We see dispersion-type and absorption-type spectra in the $m_{iir}$ and $m_{iii}$ components, respectively.
These features are well explained by a Landau-Lifshitz-Gilbert (LLG) equation
\begin{eqnarray}
\label{eq:LLG}
\frac{d{\bf M}}{dt} = -\gamma\mu_{0}({\bf M} \times {\bf H}_{eff}) + \frac{\alpha}{M_{s}}({\bf M} \times \frac{d{\bf M}}{dt}),
\end{eqnarray}
where ${\bf H}_{eff}$ is the total effective field, $\gamma$ is the gyromagnetic ratio, $\mu_{0}$ is the magnetic permeability of free space, and $\alpha$ is the Gilbert damping constant.
By solving the LLG equation~\cite{Stancil:2009ux}, we obtain $\mbox{\boldmath $\chi$}({\bf k}, \omega) = \chi({\bf k}, \omega) + i\kappa({\bf k}, \omega)$ with 
\begin{eqnarray}
\label{eq:LLGR}
\chi({\bf k}, \omega) = \frac{\omega_{0}({\bf k}) \pm \omega}{(\omega_{0}({\bf k}) \pm \omega)^{2} + \alpha^{2}\omega^{2}}
\end{eqnarray}
and 
\begin{eqnarray}
\label{eq:LLGI}
\kappa({\bf k}, \omega) = \frac{\alpha \omega}{(\omega_{0}({\bf k}) \pm \omega)^{2} + \alpha^{2}\omega^{2}},
\end{eqnarray}
where $\omega_{0}({\bf k})$ represents the dispersion relation of spin waves.
As shown by the dashed curves in Figs.~\ref{SWaTAt560Oe}(d) and~\ref{SWaTAt560Oe}(e), the $\chi$ and $\kappa$ reproduce the observed $m_{iir}$ and $m_{iii}$ components of the PSWaT spectra, respectively, with the parameters of $\omega_{0}({\bf k})/2\pi$ = 2.56 GHz and $\alpha$ = 0.03.
In the present case, the maximum time delay between the pump and probe pulses ($T_{max}$ = 13 ns) is much shorter than the spin relaxation time of garnet films so that $\alpha$ is limited by $T_{max}$ giving $\alpha \sim $1/$(\omega T_{max}) = 0.03$.


Finally, it is important to emphasize that spin waves generated through various processes with different spatial symmetries are decomposed in the different components of the PSWaT spectra. This characteristic of PSWaT makes PSWaT a powerful method for the investigation of the excitation mechanism of spin waves.

In summary, we have developed phase-resolved spin-wave tomography (PSWaT) to obtain dispersion relations of spin waves with their phase information.
The PSWaT is based on a convolution theory, a Fourier transform, and a time-resolved magneto-optical imaging method.
The phase information of spin waves is obtained by separating the real and the imaginary components of the complex Fourier transform in the PSWaT analysis~\cite{Hashimoto:2017jb}.
PSWaT was examined with a Bi-doped in-plane magnetized garnet film.
The PSWaT spectra for spin waves generated by photo-induced demagnetization are well explained by a model calculation based on the LLG equation.

\begin{acknowledgments}
We thank Dr. R. Iguchi for fruitful discussions.
This work was financially supported by JST-ERATO Grant Number JPMJER1402, and World Premier International Research Center Initiative (WPI), all from MEXT, Japan.
\end{acknowledgments}


%

\end{document}